\documentclass[pr1,amsmath,twocolumn,showkeys,superscriptaddress,preprintnumbers]{revtex4}

\usepackage{gensymb}
\usepackage{graphicx,color}
\usepackage{amssymb}
\usepackage{enumerate}
\usepackage{verbatim}
\usepackage{natbib}
\usepackage{amsmath}
\usepackage{textcomp}

\graphicspath{{./}{figs/}}

\begin{document}
  \newcommand {\nc} {\newcommand}
  \nc {\Sec} [1] {Sec.~\ref{#1}}
  \nc {\IR} [1] {\textcolor{red}{#1}} 
  \nc {\IB} [1] {\textcolor{blue}{#1}}

\title{Nuclear masses learned from a probabilistic neural network}

\author{A.~E.~Lovell}
\email{lovell@lanl.gov}
\affiliation{Theoretical Division, Los Alamos National Laboratory, Los Alamos, NM, 87545, USA}
\author{A.~T.~Mohan}
\affiliation{Computational Division, Los Alamos National Laboratory, Los Alamos, NM, 87545, USA}
\author{T.~M.~Sprouse}
\affiliation{Theoretical Division, Los Alamos National Laboratory, Los Alamos, NM, 87545, USA}
\author{M.~R.~Mumpower}
\affiliation{Theoretical Division, Los Alamos National Laboratory, Los Alamos, NM, 87545, USA}

\date{\today}
\preprint{LA-UR-21-27783}


\begin{abstract}
Machine learning methods and uncertainty quantification have been gaining interest throughout the last several years in low-energy nuclear physics.  In particular, Gaussian processes and Bayesian Neural Networks have increasingly been applied to improve mass model predictions while providing well-quantified uncertainties.  In this work, we use the probabilistic Mixture Density Network (MDN) to directly predict the mass excess of the 2016 Atomic Mass Evaluation within the range of measured data, and we extrapolate the inferred models beyond available experimental data. The MDN not only provides mean values but also full posterior distributions both within the training set and extrapolated testing set.  We show that the addition of physical information to the feature space increases the accuracy of the match to the training data as well as provides for more physically meaningful extrapolations beyond the the limits of experimental data. 
\end{abstract}


\maketitle

\section{Introduction}
\label{sec:intro}


The mass of a nucleus is one of the most basic properties that it possesses. 
Predicting this key quantity for the many thousands of atomic nuclei that exist in nature represents a major hurdle for theoretical nuclear physics modeling. 
Many applications such as those in astrophysics~\cite{Mumpower2016r, Horowitz2019, Vassh2021, Cowan2021}, reaction modeling~\cite{Kawano2006, Kawano2016}, and nuclear data evaluations~\cite{Chadwick2011} are sensitive to the accurate description of nuclear masses. 

Several theoretical approaches afford the prediction of nuclear masses. 
Phenomenological models like the liquid drop model combine experimental observations with physically interpreted parameters and are successful in predicting bulk properties across the nuclear chart.  
More advanced mass formula, such as the Duflo-Zuker model, incorporate microscopic effects via analytic expressions~\cite{Duflo1995, Liu2011}. 
A more detailed picture can be provided by macroscopic-microscopic models which tack on a microscopic component to the bulk liquid drop terms, in the form of shell and pairing corrections~\cite{Moller2016}. 
Such modeling has proven to be robust in predicting the properties of atomic nuclei, offering the best accuracy of masses at the present time~\cite{Koura2005, Zhang2014}. 
Recent advances in computation open the opportunity for a more microscopic picture of the nucleus, as with the approach of using energy density functionals which are capable of prediction many properties beyond masses~\cite{Goriely2009, Schunck2016}. 
At this time, to achieve accuracy on the order of macroscopic-microscopic models, fully microscopic models rely on the addition of phenomenological parameters~\cite{Goriely2001, Goriely2013b, Goriely2014}. 

Over the past several years, quantifying uncertainties on nuclear models has become a focus of the community, \emph{e.g.}~\cite{Phillips2021}.  There has been a large push to understand uncertainties coming from parameter optimization and model truncation in nuclear theory, from \emph{ab initio} methods~\cite{Holt2019,Konig2020} to few-body methods~\cite{Lovell2015,Lovell2017,King2018,Lovell2018,King2019} and across Effective Field Theories~\cite{Schindler2009,Perez2015,Furnstahl2015,Furnstahl2015a,Wesolowski2016,Melendez2017} and Density Functional Theories~\cite{Schunck2015,McDonnell2015}.  Much of the focus has been in moving the field from standard $\chi^2$ minimization and covariance propagation to Bayesian methods.  However, there has also been significant effort put into emulators such as Gaussian Processes (GP)~\cite{Novak2014,Sangaline2016}, Bayesian Neural Networks (BNN)~\cite{Utama2016,Utama2016a,Wang2019}, a committee of Neural Networks~\cite{Lasseri2020}, and most recently Eigenvector Continuation~\cite{Frame2018,Sarkar2021,Koenig2020,Ekstrom2019,Wesolowski2021,Furnstahl2020} for speeding up parameter optimization, improving the accuracy of nuclear models, and uncertainty quantification.  In particular, there has been significant developments surrounding the application of GPs and BNNs to improving nuclear mass predictions~\cite{Neufcourt2018,Neufcourt2019,Neufcourt2020,Neufcourt2020a}.

Model calibration, or the fitting of model parameters to data, is an important aspect of this type of research.
In~\cite{Neufcourt2018,Neufcourt2019,Neufcourt2020,Neufcourt2020a}, the GP and BNN emulators were trained on the differences between the Atomic Mass Evaluation and various mass models (the residuals).  These trained residuals were used to `correct' each of the mass models to lie within the range of the Atomic Mass Evaluation, as well provide uncertainties on the theoretical predictions.  However, there are some challenges using both the BNN and GP.  Neufcourt, et. al.~\cite{Neufcourt2018} found that the GP is much more stable than the BNN, in particular when the size of the training set available for the neutral network training is limited.  In addition, they note that by using statistical arguments, the size of the neural network in the BNN cannot exceed certain dimensions due, again, to the size of the training set and is further limited by the number of free parameters given available data.  (However, some studies, e.g. \cite{Nakkiran2019}, show that as the complexity of a neural network increases, it will first perform worse and then perform better and that there are certain regimes where increasing the number of training points worsens the performance.)  The GP also has some downsides, the major one being that its predictions return to the mean value after one or two correlation lengths, so there is no straight-forward way to extrapolate the residual corrections.  Although it is worthwhile to note that one must be careful any time that an extrapolation is performed, regardless of the machine learning method used.

In this work, we take a different approach than either the GP or BNN and instead, use a probabilistic machine learning technique, the Mixture Density Network (MDN)~\cite{MDN}.  Unlike most neural networks which directly map the input feature space to a deterministic output space, the MDN describes the output as a mixture of Gaussian functions and uses the neural network to learn the weights, means, and standard deviations of these distributions.  The Gaussian distribution gives not only mean values but also uncertainties associated with each prediction, and the mixture allows full posterior distributions to be predicted.  This method allows us to take into account discrepant values in the training set without having to simply average them.  We explored the ability of the MDN to reliably propagate uncertainties on fission fragment mass yields from the training set to predictions in~\cite{Lovell2020} for both interpolations within the training set and extrapolations beyond it.  The uncertainties were found to reliably propagate from the training set to the predictions.  It was found that very small experimental uncertainties became about 10\% larger in the predictions compared to the training set, but this deviation was caused mainly by the several orders of magnitude that the fission yields span.  This challenge is completely mitigated in the current work because the mass excesses do not cover multiple orders of magnitude.  


This paper is divided into the following sections:  in Sec. \ref{sec:methods}, we discuss the details of the MDN, the training and testing sets, as well as the neural network architecture; in Sec. \ref{sec:results}, we discuss the results of the various training sets that we investigated and how adding more physics information into the feature space improves the training and extrapolation; finally, we conclude and describe some future directions for this work in Sec. \ref{sec:conclusion}.


\section{Theory}
\label{sec:methods}

In this section, we describe our choice of probabilistic neural network, the Mixture Density Network (MDN), our training and testing sets, and the neural network architecture.

\subsection{Mixture Density Network (MDN)}

In a standard feed-forward Neural Network (NN), the goal is to optimize a complex, non-linear mapping between an input feature space $\textbf{x}$ and an output space $\textbf{y}$ though $\textbf{y} = f(\textbf{x})$.  This mapping is typically deterministic--in that stochasticity is not taken into account--and these networks tend to be incapable of well reproducing probabilistic training sets.  To mitigate these challenges, and in particular to take uncertainties on the training set into account, we employ a probabilistic machine learning technique, the Mixture Density Network~\cite{MDN}.  Instead of using the NN to directly map the input to the output of the training set, the MDN describes the output as a mixture of Gaussian functions,
\begin{equation}
\textbf{y}(\textbf{x}) = \sum \limits _{i=1} ^m \alpha _i (\textbf{x}) \mathcal{N} \left [ \mu _i (\textbf{x}), \sigma _i (\textbf{x}) \right ],
\label{eqn:MDNGauss}
\end{equation}

\noindent where $\mathcal{N}$ is the normal distribution and the weights, means, and standard deviations, $\alpha_i (\textbf{x})$, $\mu _i (\textbf{x})$, $\sigma _i (\textbf{x})$, are learned by the standard feed-forward NN.  The loss function changes from a simple mean-squares estimate to a log-likelihood loss,
\begin{equation}
\mathcal{L} = - \ln {\left [ \sum \limits _{i=1} ^m \frac{\alpha_i(\textbf{x})}{(2\pi)^{m/2} \sigma_i (\textbf{x})} \mathrm{exp} \left \{ -\frac{|| \textbf{t}-\mu_i(\textbf{x}) ||^2}{2\sigma_i (\textbf{x})^2} \right \} \right ]},
\label{eqn:logloss}
\end{equation} 

\noindent where $\textbf{t}$ is the vector of training outputs and $m$ is the total number of Gaussian mixtures.  Instead of minimizing the absolute root mean square error, the loss function of Eq. (\ref{eqn:logloss}) minimizes the difference in the probability distributions between the true distribution of the training set and the predicted posterior distribution.

To better represent probabilistic data sets than traditional deterministic neural networks, the MDN has the additional benefit of predicting the exact posterior distribution of each predicted value instead of simply predicting a mean and a standard deviation--then leaving it to the user to make a decision on the shape of the posterior distribution.  The mixture of Gaussians also takes into account discrepant points in the training set, one such example shown in~\cite{Lovell2020}, without just taking a weighted average, as the MDN has the flexibility to form a joint distribution over the discrepant sets.  More details about the MDN can be found in Ref.~\cite{MDN}.

Our MDN is written in \texttt{PyTorch}~\cite{pytorch} and can be run on both CPU and GPU, with a significant speed up when using the GPU implementation due to the parallel nature of the computational workload.  The hyperparameters that set the number of Gaussian mixtures in the MDN and the number of layers and nodes in the feed-forward neural network can be specified for each run without any change in the source code.  In addition, the implementation is generalized to allow for any number of features in the input vector and for the prediction of any number of outputs, which is explored in this work.  In the current implementation, the number of Gaussian mixtures is constrained to be the same for each output, however, this could be easily generalized.

\subsection{The training and testing sets}
\label{sec:models}

While the goal of this work is to directly predict the mass excess, $\delta M$, instead of predicting a correction to any one specific mass model, we primarily focus on how the features included in the input space lead to better predictions within the training set and when extrapolating beyond it.  To that end, there are several different training sets that we define for this work, which we will describe in this section, each successive one adding more information about the nuclei with the goal to decrease the discrepancy between the predicted values and the experimentally evaluated mass excesses.  In addition, we also explore how constraining multiple outputs improves the training, particularly predicting both the mass excess and one-neutron separation energy, $S_n$.  To clearly denote these distinctions, we indicate mass predictions only with \emph{M}, both mass and separation energy predictions with \emph{MS}, and the number that follows indicates the length of the feature space.

In our base model, \emph{M2}, the input feature space consists of just the number of neutrons, $N$, and the number of protons, $Z$, and the only we are only predicting the mass excess.  
The inputs for model \emph{M6} take inspiration from the standard semi-empirical mass model~\cite{Weizsacker1935}.  In addition to $N$ and $Z$, we include other bulk properties in the feature space:  the mass number, $A$, and $A^{2/3}$ (from the volume and surface terms), $Z(Z-1)/A^{1/3}$ from the Coulomb term, and $(N-Z)^2/A$ from the asymmetry term.  In addition, \emph{M8} also includes pairing information, $Z_{EO}$ and $N_{EO}$ where $Z_{EO}$ ($N_{EO}$) is 0 if $Z$ ($N$) is even and 1 if $Z$ ($N$) is odd.  Then \emph{M10} and \emph{M12} include information about magic numbers and the nuclear shells.  \emph{M10} adds $\Delta N$ and $\Delta Z$ to the features of \emph{M8}, which are defined as the number of neutrons or protons on top of a magic number or away from a magic number (whichever is smaller), \emph{i.e.} the valance nucleons.  For \emph{M12}, $N_\mathrm{shell}$ and $Z_\mathrm{shell}$ are the shell model orbitals of the last neutron and proton, starting counting at 0, as defined in \emph{e.g.}~\cite{ReactionsBook}.  We additionally construct an equivalent \emph{MS} training set for each feature space.  An overview of of the MDN models can be found in Table~\ref{tab:trainingSets}.

\begin{table}
\centering
\begin{tabular}{ccc}
\hline \hline \textbf{Model Name} & \textbf{Feature Space} & \textbf{Output} \\ \hline
M2 & $N$, $Z$ & $\delta M$ \\ 
M6 & $N$, $Z$, $A$, $A^{2/3}$,  & $\delta M$\\
& $Z(Z-1)/A^{1/3}$, $(N-Z)^2/A$ & \\
M8 & $N$, $Z$, $A$, $A^{2/3}$, $Z(Z-1)/A^{1/3}$, & $\delta M$ \\
&  $(N-Z)^2/A$, $Z_{EO}$, $N_{EO}$ & \\
M10 & $N$, $Z$, $A$, $A^{2/3}$, $Z(Z-1)/A^{1/3}$, & $\delta M$ \\
&  $(N-Z)^2/A$, $Z_{EO}$, $N_{EO}$ & \\
& $\Delta N$, $\Delta Z$ & \\
M12 & $N$, $Z$, $A$, $A^{2/3}$, $Z(Z-1)/A^{1/3}$, & $\delta M$ \\
&  $(N-Z)^2/A$, $Z_{EO}$, $N_{EO}$ & \\
& $\Delta N$, $\Delta Z$, $N_\mathrm{shell}$, $Z_\mathrm{shell}$ & \\
MS2 & $N$, $Z$ & $\delta M$, $S_n$ \\
MS6 & $N$, $Z$, $A$, $A^{2/3}$,  & $\delta M$, $S_n$ \\
& $Z(Z-1)/A^{1/3}$, $(N-Z)^2/A$ & \\
MS8 & $N$, $Z$, $A$, $A^{2/3}$, $Z(Z-1)/A^{1/3}$, & $\delta M$, $S_n$ \\
&  $(N-Z)^2/A$, $Z_{EO}$, $N_{EO}$ & \\
MS10 & $N$, $Z$, $A$, $A^{2/3}$, $Z(Z-1)/A^{1/3}$, & $\delta M$, $S_n$ \\
&  $(N-Z)^2/A$, $Z_{EO}$, $N_{EO}$ & \\
& $\Delta N$, $\Delta Z$ & \\
MS12 & $N$, $Z$, $A$, $A^{2/3}$, $Z(Z-1)/A^{1/3}$, & $\delta M$, $S_n$ \\
&  $(N-Z)^2/A$, $Z_{EO}$, $N_{EO}$ & \\
& $\Delta N$, $\Delta Z$, $N_\mathrm{shell}$, $Z_\mathrm{shell}$ & \\
\hline \hline 
\end{tabular}
\caption{Description of the training sets used in this work.  Additional details can be found in the text.}
\label{tab:trainingSets}
\end{table}

We construct our training sets from the 2016 Atomic Mass Evaluation (AME2016).  For each isotope where the separation energy can be defined (e.g. where the isotope with N-1 neutrons has also been measured, 2074 nuclei), 25 values of $\delta M$ are sampled from a normal distribution, $\mathcal{N}(\delta M(N,Z), \Delta M(N,Z))$, where $\Delta M (N,Z)$ is the experimentally evaluated uncertainty on $\delta M (N,Z)$.  The same is used to calculate $S_n(N,Z)$ except that the masses are sampled from both $\mathcal{N}(\delta M(N,Z), \Delta M(N,Z))$ and $\mathcal{N}(\delta M(N-1,Z), \Delta M(N-1,Z))$ with
\begin{equation}
S_n(N,Z) = \Delta _n - \delta M (N,Z) + \delta M (N-1,Z),
\label{eqn:Sn}
\end{equation}

\noindent where $\Delta _n = 8.0713$ MeV/$c^2$ is the neutron mass excess.  Sampling from the distribution on both nuclei allows us to include an uncertainty on the one-neutron separation energy that comes from the evaluated uncertainty on both masses that are needed to calculate this quantity.

The testing set consists of all of the nuclei predicted by the 2012 version of the Finite-Range Droplet Model (FRDM)~\cite{Moller2016}, 8668 nuclei.  This set of nuclei was chosen to span a large range in $N$ and $Z$, beyond the experimentally measured nuclei.  The FRDM predictions in unmeasured regions of the chart have been consistent over the years, and therefore, a comparison of the MDN predictions to FRDM is representative of the quality of the extrapolation.  


Training ensues with the minimization of the loss function defined in Eq. (\ref{eqn:logloss}).  After training, we quantify the goodness of fit in the standard way by calculating the average standard deviation between the MDN prediction and the quantity of interest through the $\sigma_\mathrm{RMS}$.  For example, $\sigma_\mathrm{RMS}$ between the predicted MDN mass excess, $\delta M ^\mathrm{MDN}$, and the AME mass excess, $\delta M ^\mathrm{AME}$, is defined as 
\begin{equation}
\sigma_\mathrm{RMS} = \sqrt{\frac{\sum \limits _i (\delta M_i ^\mathrm{AME} -\delta M_i ^\mathrm{MDN})^2}{K}},
\label{eqn:sigmaRMS}
\end{equation}

\noindent where $K$ is the total number of nuclei that were included in the prediction.  The same quantity can also be defined for $S_n$. 

\subsection{Network architecture}

In this work, we are investigating how to improve network predictions by including more physics in the input feature space.  Although other studies, e.g. \cite{Wang2021}, have shown that deeper networks are better able to capture complexities in the data sets of interest compared to NNs with a single layer, deeper networks do not \emph{a priori} provide any insight into why a model is or is not training well.  More complex networks also do not lead to any new understanding of the underlying physics, even though they are exceedingly useful for building emulators.  To that purpose, we have constructed the neural network to be as small as possible while still providing very good agreement between the training set and the MDN predictions (what is meant by good agreement will be discussed as the results are presented).  To that end, our network consists of a single hidden layer with only six nodes.  As in previous work~\cite{Lovell2020}, because the training set was pulled from a single Gaussian distribution at each mass, our MDN consists of only a single Gaussian mixture.  The MDN predictions do not change significantly when multiple Gaussians are used--up to some point when the training becomes unstable.  When multiple Gaussians are used, only one Gaussian has a non-zero weight for each mass, which further supports our decision to include only a single Gaussian in our MDN. 

Our MDN in $\texttt{PyTorch}$ uses an Adam optimizer with a learning rate of 0.001 (the default), with a hyperbolic tangent as the activation function.  The weights of the NN are initialized randomly.  The MDN was found to be converged after training for 100,000 epochs, which is used throughout this work.


\section{Results and Discussion}
\label{sec:results}


\subsection{Feature space engineering}

\begin{figure*}
\centering
\begin{tabular}{cc}
\includegraphics[width=0.5\textwidth]{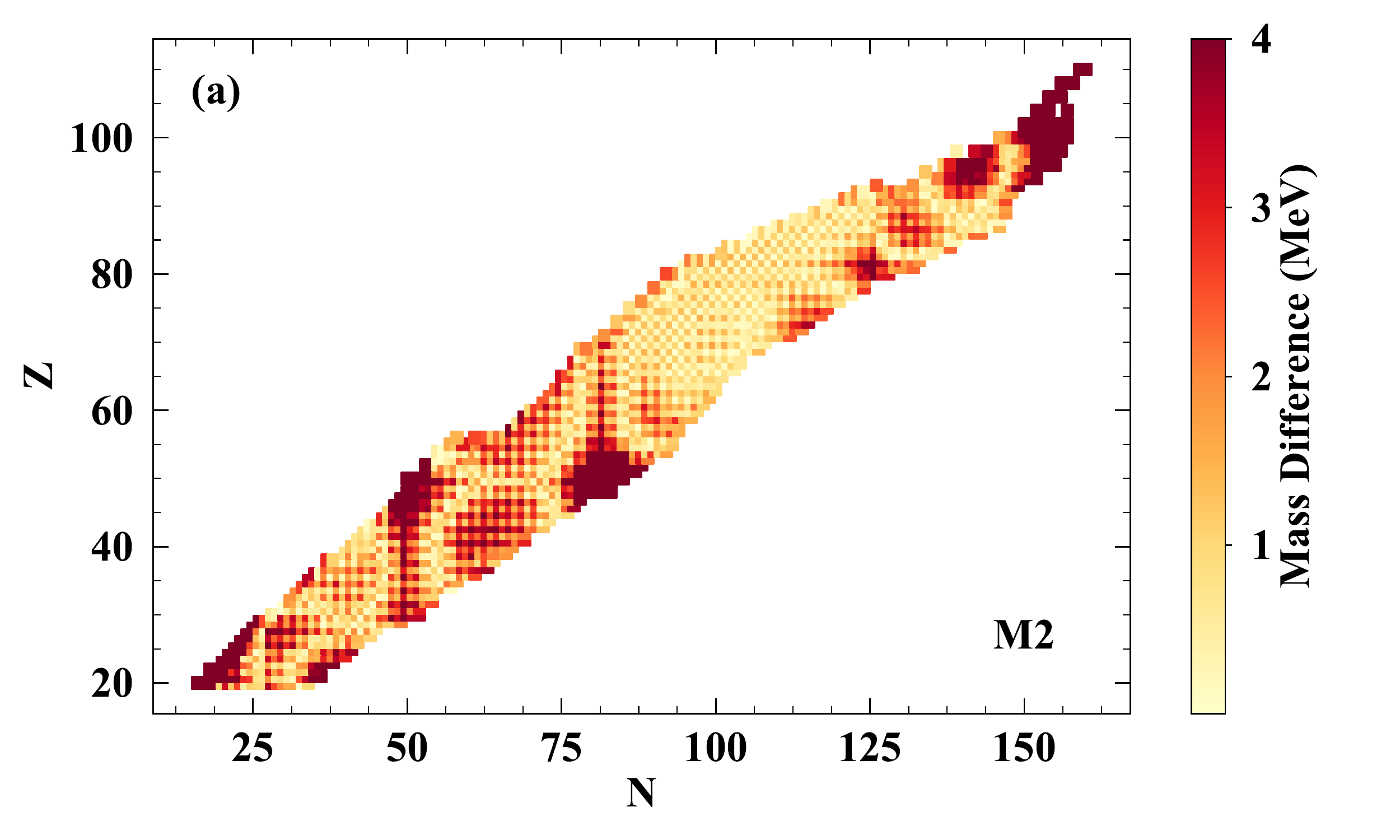} & \includegraphics[width=0.5\textwidth]{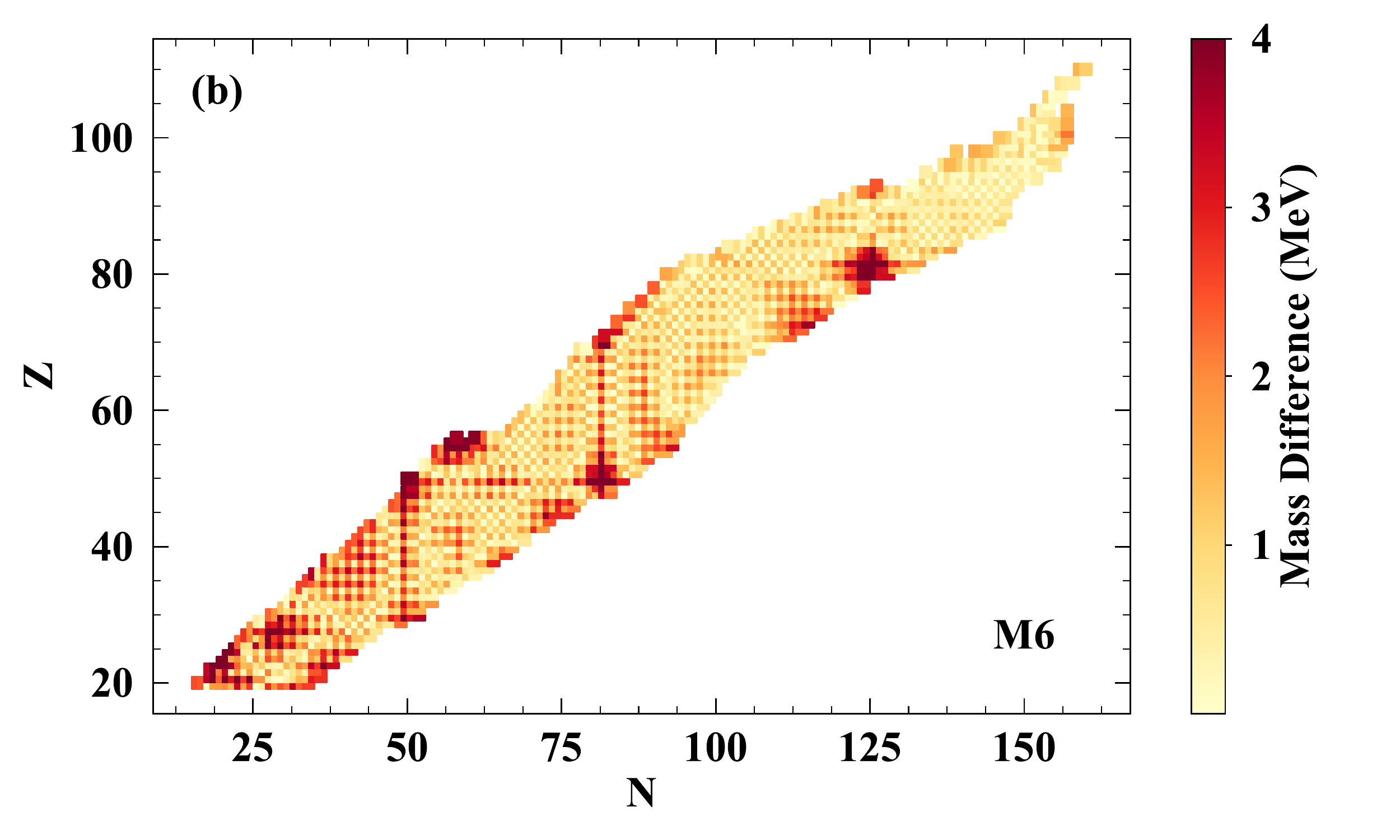} \\
\includegraphics[width=0.5\textwidth]{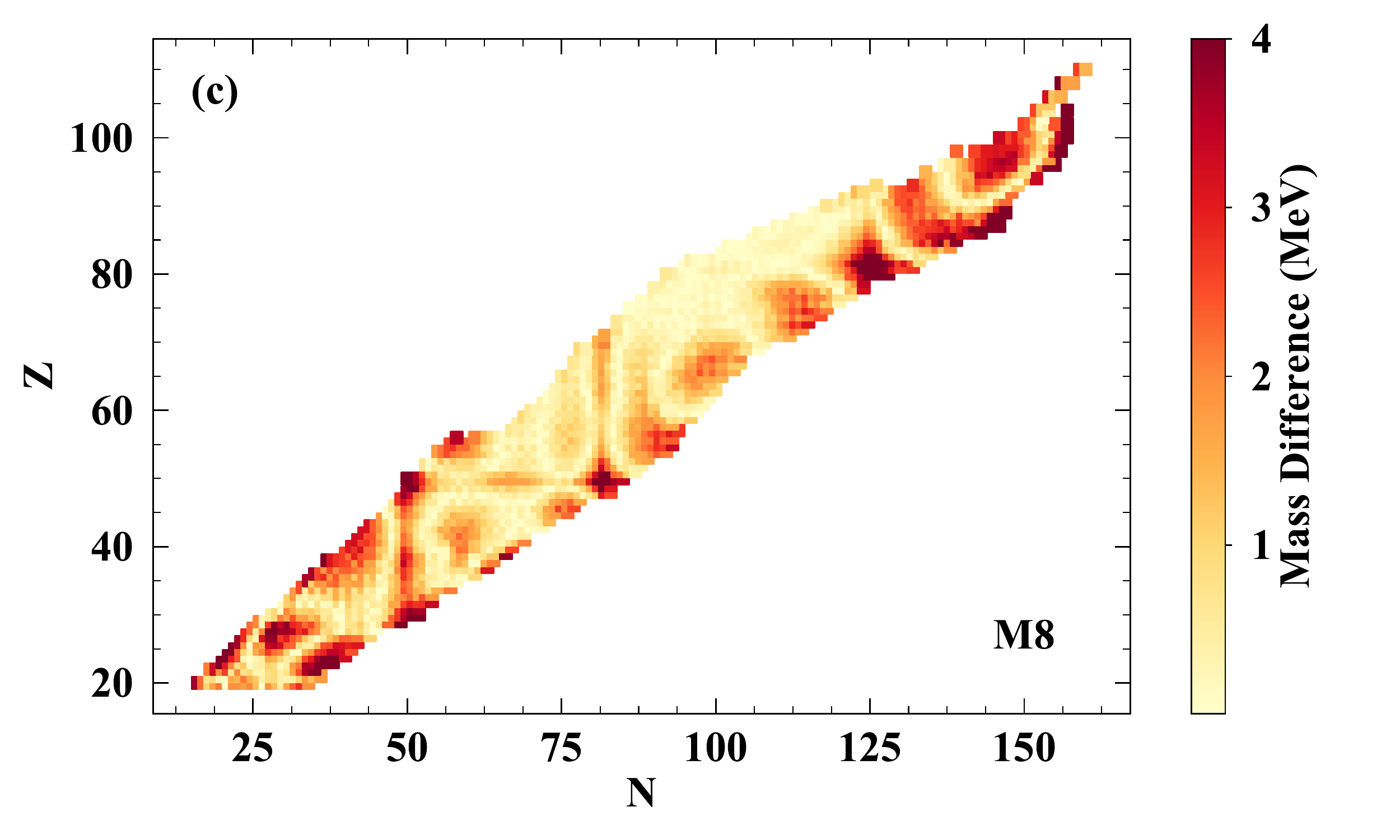} & \includegraphics[width=0.5\textwidth]{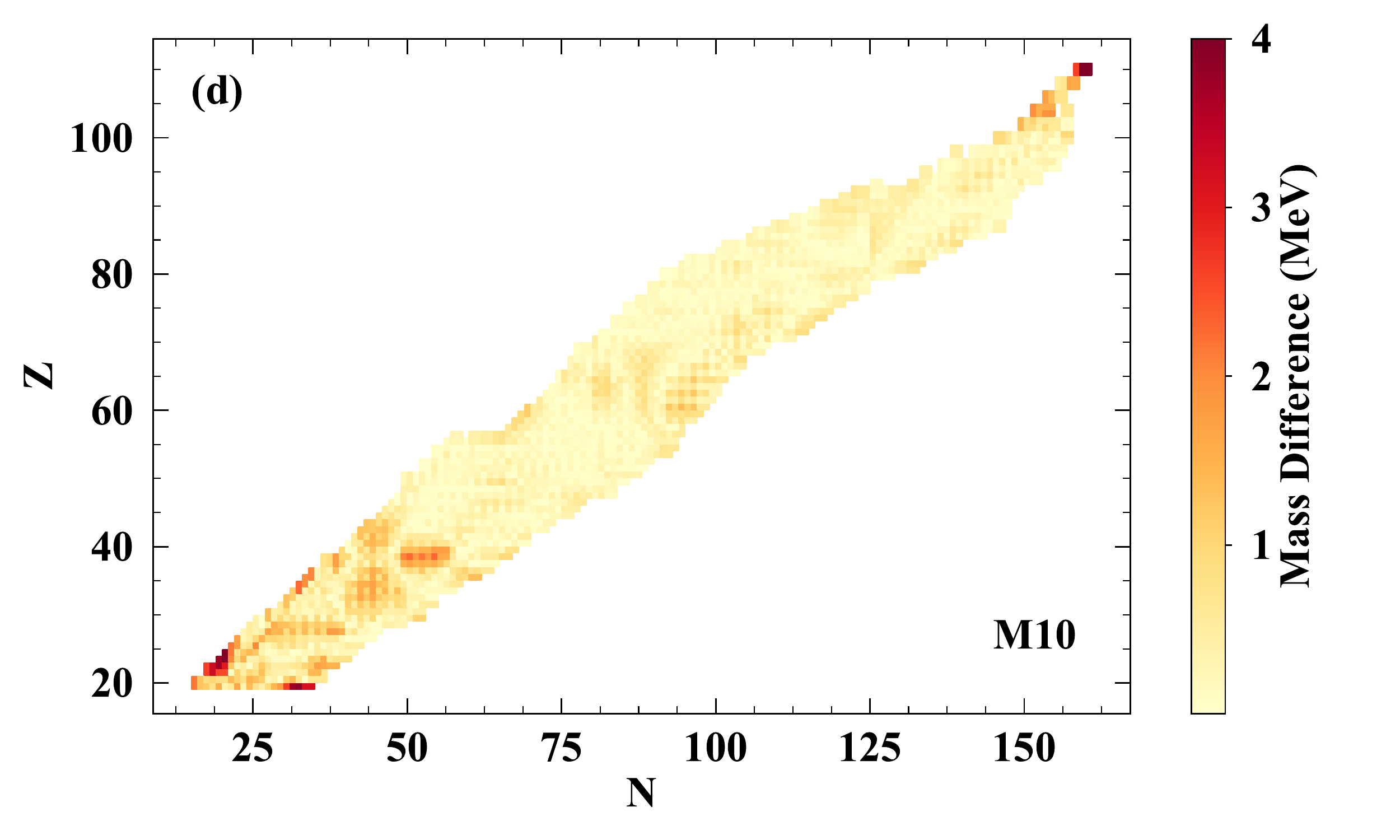} \\
\includegraphics[width=0.5\textwidth]{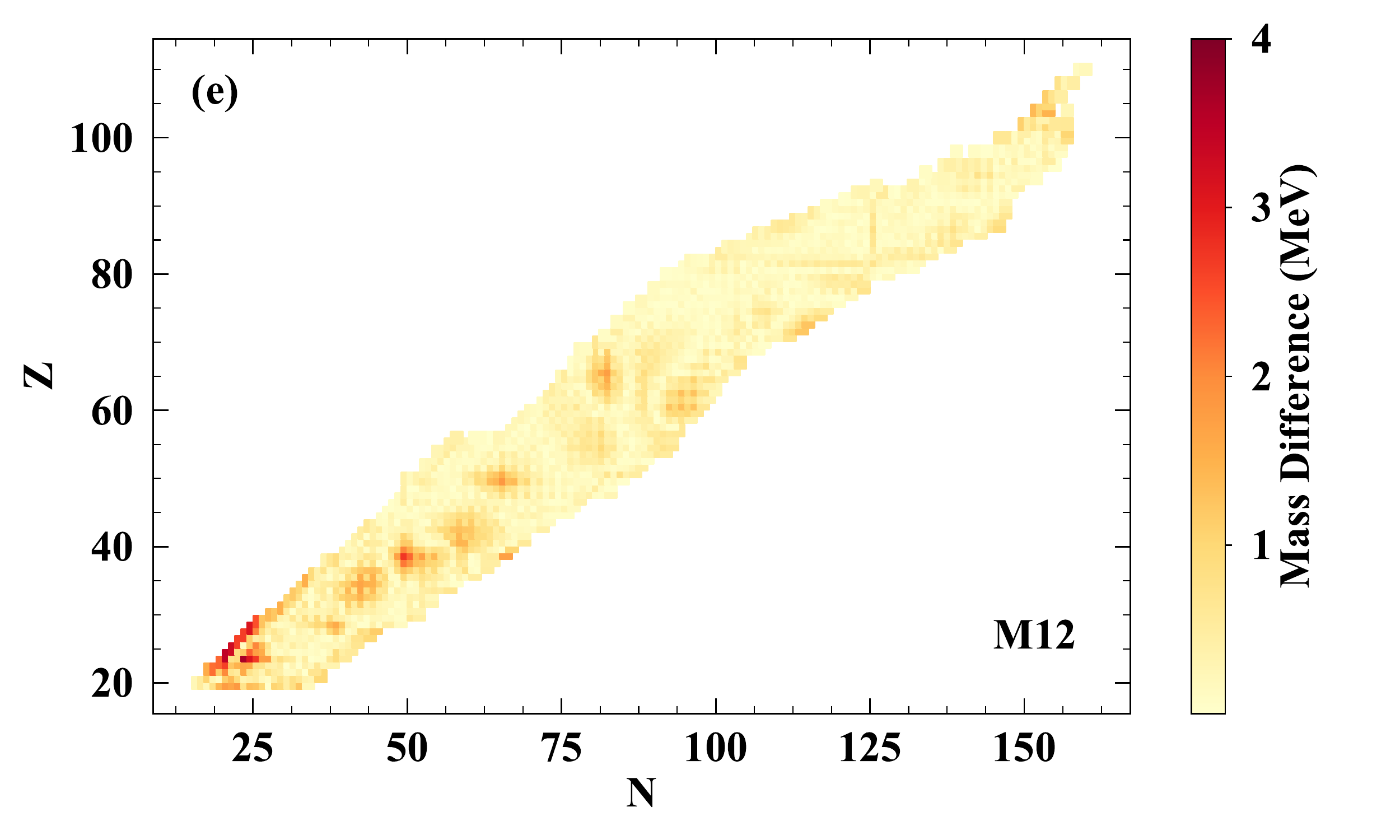} & \includegraphics[width=0.5\textwidth]{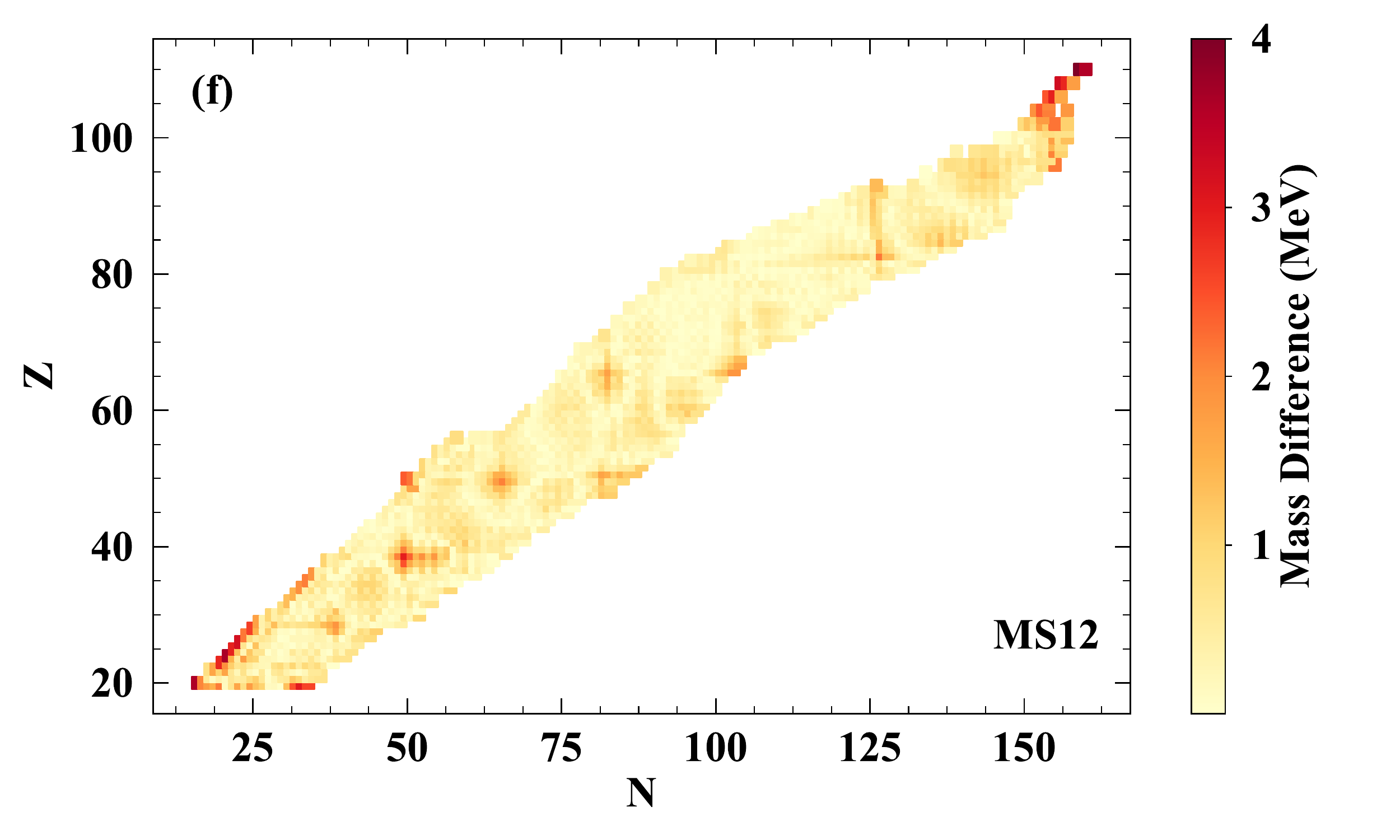} \\
\end{tabular}
\caption{Mass difference between the MDN predictions and the AME2016 (in MeV) across the nuclear chart for the six models studied in this work, (a) \emph{M2}, (b) \emph{MS2}, (c) \emph{MS6}, (d) \emph{MS8}, (e) \emph{MS10}, and (f) \emph{MS12}.  Darker colors indicate regions of larger differences between the AME and the MDN predictions.}
\label{fig:massDifference}
\end{figure*}

Figure \ref{fig:massDifference} shows an overview of of the absolute value of the mass difference between the AME2016 masses and the output of the MDN for most of the models in Table \ref{tab:trainingSets} (note that the only \emph{MS} model that we show is \emph{MS12}, as the residuals of each of the $MN$ and $MSN$ show similar features).  Beginning with panel (a) for model \emph{M2}, which only includes the number of neutrons and protons in the input feature space and trains on just the mass excess, we immediately see that such a small network cannot adequately reproduce the training set, particularly not to the level of accuracy that most mass models reach.  This is quantified in the first row of Table \ref{tab:ResultsOverview} which lists the $\sigma_\mathrm{RMS}$ of the MDN predictions with respect to the AME2016.  For model \emph{M2}, $\sigma_\mathrm{RMS}=3.90$ MeV where as most mass models with phenomenological corrections are on the order of 0.5 MeV, e.g.~the HFB approach~\cite{Goriely2013}. 

The clearest trend that is seen in the \emph{M2} model is that the difference in the masses increase as the the nuclei become heavier--most noticeably in the highest-Z region of the nuclear chart.  This trend leads us to model \emph{M6}, which introduces several factors from the semi-empirical mass model into the feature space.  Although we could improve the training by tuning the hyperparameters of the neural network (including more layers and nodes), this only serves to lower $\sigma_{RMS}$ and does not improve the extrapolation beyond the training set or the consistency of the extrapolations (discussed later in Sections \ref{sec:extrapolation} and \ref{sec:robustness}).  Instead, by including these four values into the feature space, $\sigma_\mathrm{RMS}$ is reduced to 1.57 MeV, a decrease of a factor of almost 3.  The residuals between the AME2016 masses and the masses predicted by \emph{M6} are shown in Fig. \ref{fig:massDifference}(b).  Now, as the residuals have drastically decreased in the high-Z region of the nuclear chart, it is much easier to see patterns in the residuals.  The residuals are noticeably larger at the magic numbers, a prominent feature seen in the mass models studied in~\cite{Neufcourt2018} for the two-neutron separation energies.  In addition, there is clear odd-even staggering seen in the residuals, meaning that the MDN predictions do not reproduce the staggering that is observed in the data.


\begin{table}
\centering
\begin{tabular}{ccc}
\hline \hline
\textbf{Model} & \textbf{$\delta M$ $\sigma_\mathrm{RMS}$ (MeV)} & \textbf{$S_n$ $\sigma_\mathrm{RMS}$ (MeV)} \\ \hline
M2 & 3.90 & --- \\
MS2 & 2.43 & 1.25 \\
M6 & 1.57 & --- \\
MS6 & 2.07 & 1.21 \\
M8 & 1.66 & --- \\
MS8 & 2.21 & 0.57 \\
M10 & 0.58 & --- \\
MS10 & 0.76 & 0.57 \\
M12 & 0.56 & --- \\ 
MS12 & 0.64 & 0.47 \\ 
\hline \hline
\end{tabular}
\caption{$\sigma_\mathrm{RMS}$ for the mass excess (column two) and one-neutron separation energy (column three) for each of the models described in this work (column one).}
\label{tab:ResultsOverview}
\end{table}

We first take into account the even-odd staggering.  For each nucleus, we include in the feature space $N_{EO}$ and $Z_{EO}$, as described in Sec. \ref{sec:models}.  The mass residuals when this feature space is trained on, \emph{M8}, are shown in Fig. \ref{fig:massDifference}(c).  Although there is not an overall decrease in $\sigma_\mathrm{RMS}$--it is now 1.66 MeV, which is actually a slight increase from \emph{M6}--the mass residuals no longer show the odd-even staggering that is seen in Fig. \ref{fig:massDifference}(b).  We then can take into account the shell closures, by including in the feature space the distance to the shell closure, as in model \emph{M10}.  As can be seen in Fig. \ref{fig:massDifference}(d), the large differences between the AME2016 masses and the MDN predictions at the magic numbers are completely removed.  In addition, $\sigma_\mathrm{RMS}$ again decreases by almost a factor of 3, to 0.58 MeV which is on par with theoretical nuclear mass models. 

The last two models shown, \emph{M12} and \emph{MS12}, are then aimed at improving the extrapolation of the masses beyond the training set, discussed in detail in Section \ref{sec:extrapolation}.  To improve the extrapolation, we first include the shell model orbitals that each $N$ and $Z$ falls in, as described in Sec. \ref{sec:models}.  The difference between the AME2016 masses and the MDN predicted masses are shown in Fig. \ref{fig:massDifference}(e).  Within the training space, there is not much difference between models \emph{M10} and \emph{M12}, and indeed, $\sigma_\mathrm{RMS}$ only decreases by 0.02 MeV when the shells are included.  

In the last model shown, \emph{MS12}, we include in the prediction both the mass excess, $\delta M$, and the one-neutron separation energy, $S_n$.  Because $S_n$ is defined by the differences in masses, as in Eq. (\ref{eqn:Sn}), including $S_n$ in the loss function should provide an additional constraint on the slope of the mass surface, in addition to the values of the masses.  When including this additional constraint, the $\sigma_\mathrm{RMS}$ on $\delta M$ increase by $\sim 0.1$ MeV to 0.64 MeV, but the trends across the measured region of the nuclear chart remain similar to those for \emph{M12}, as shown in Fig. \ref{fig:massDifference}(f).  In addition, $\sigma_\mathrm{RMS}$ for the one-neutron separation energies of the MDN predictions compared to the AME2016 is 0.47 MeV.  Even though we have not shown the difference between models \emph{MN} and \emph{MSN} for $N=2,6,8,10$, the results shown for the difference between \emph{M12} and \emph{MS12}--where the extrapolation of the mass difference is improved with the addition of including the neutron separation energy in the output vector--are illustrative of the differences between each of these training pairs.  The extrapolation is discussed further in Sec. \ref{sec:extrapolation}.

Additionally, the addition of features to the input space reduces the mass uncertainty that is predicted by the MDN.  Particularly for model \emph{M2}, outside of the training set, the predicted uncertainties on the masses quickly become so large as to be unusable.  Naturally, the MDN predicts increased uncertainties as the nuclei become more exotic (either neutron-rich or proton-rich).  An increase in uncertainties away from the training set is exactly what we would expect from any robust prediction of uncertainties.  In addition, as more features are included in the training set, not only are the predictions within the range of the training set more accurate, but the uncertainties within the training set and extrapolated region become smaller.

\subsection{Extrapolation beyond the training set}
\label{sec:extrapolation}

\begin{figure*}
\centering
\begin{tabular}{cc}
\includegraphics[width=0.5\textwidth]{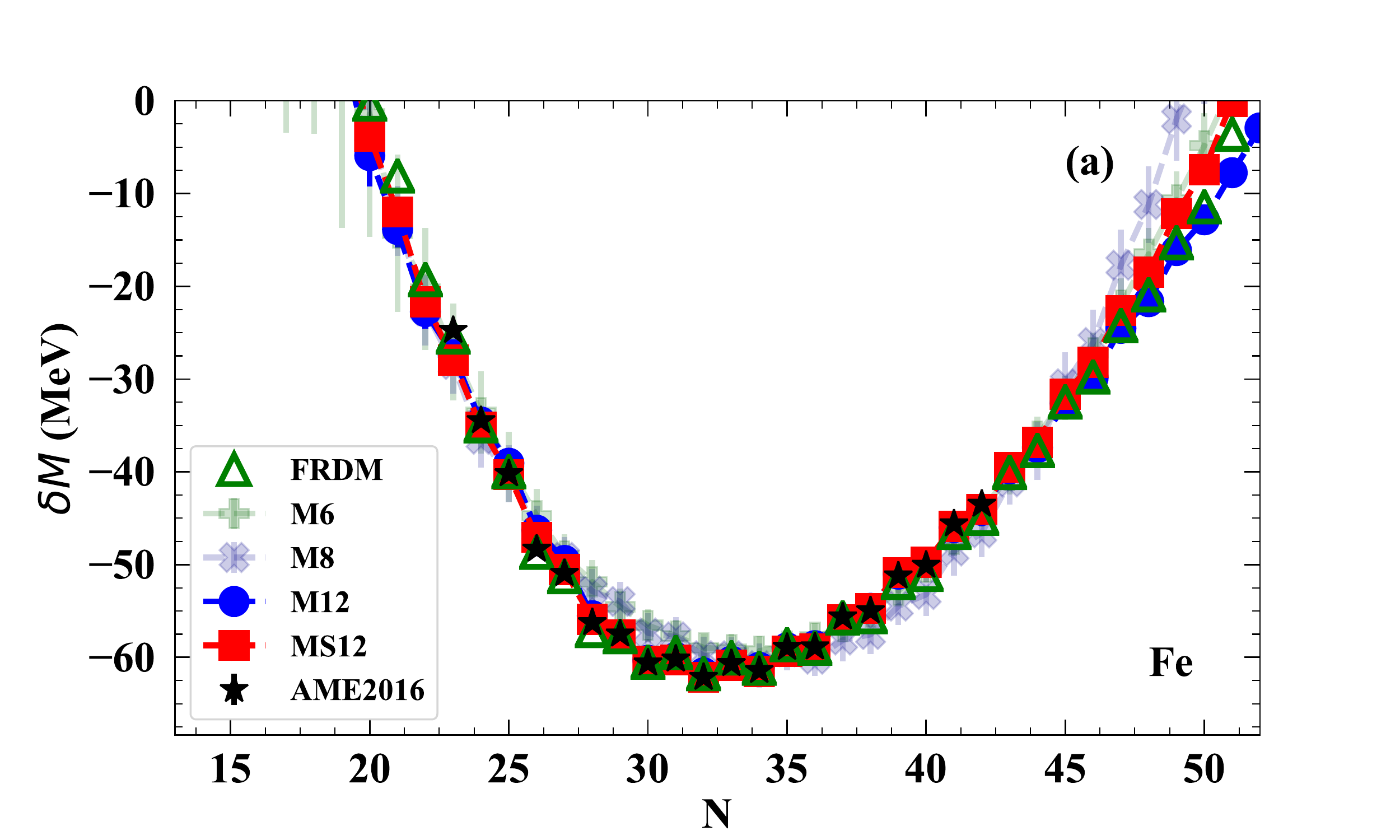} & \includegraphics[width=0.5\textwidth]{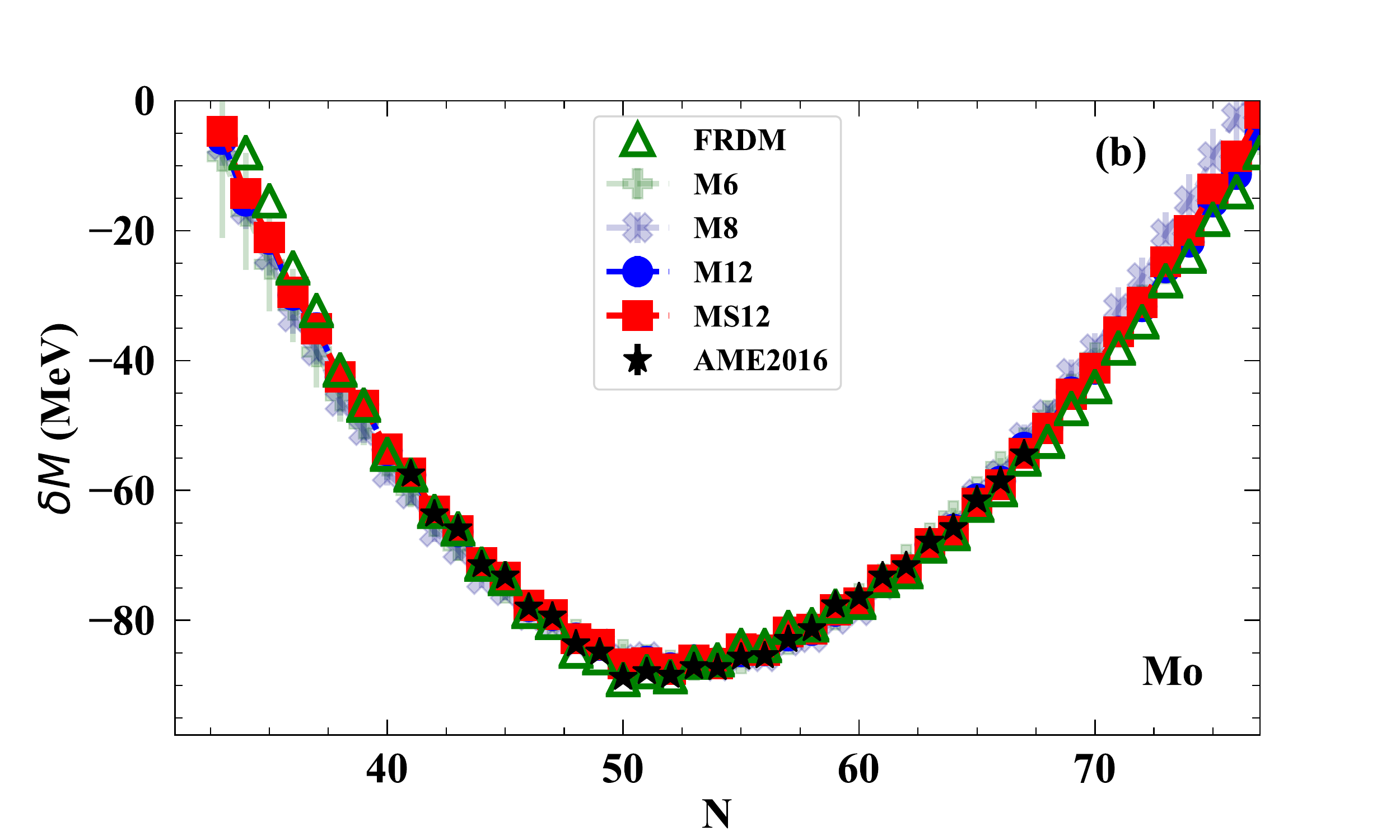} \\
\includegraphics[width=0.5\textwidth]{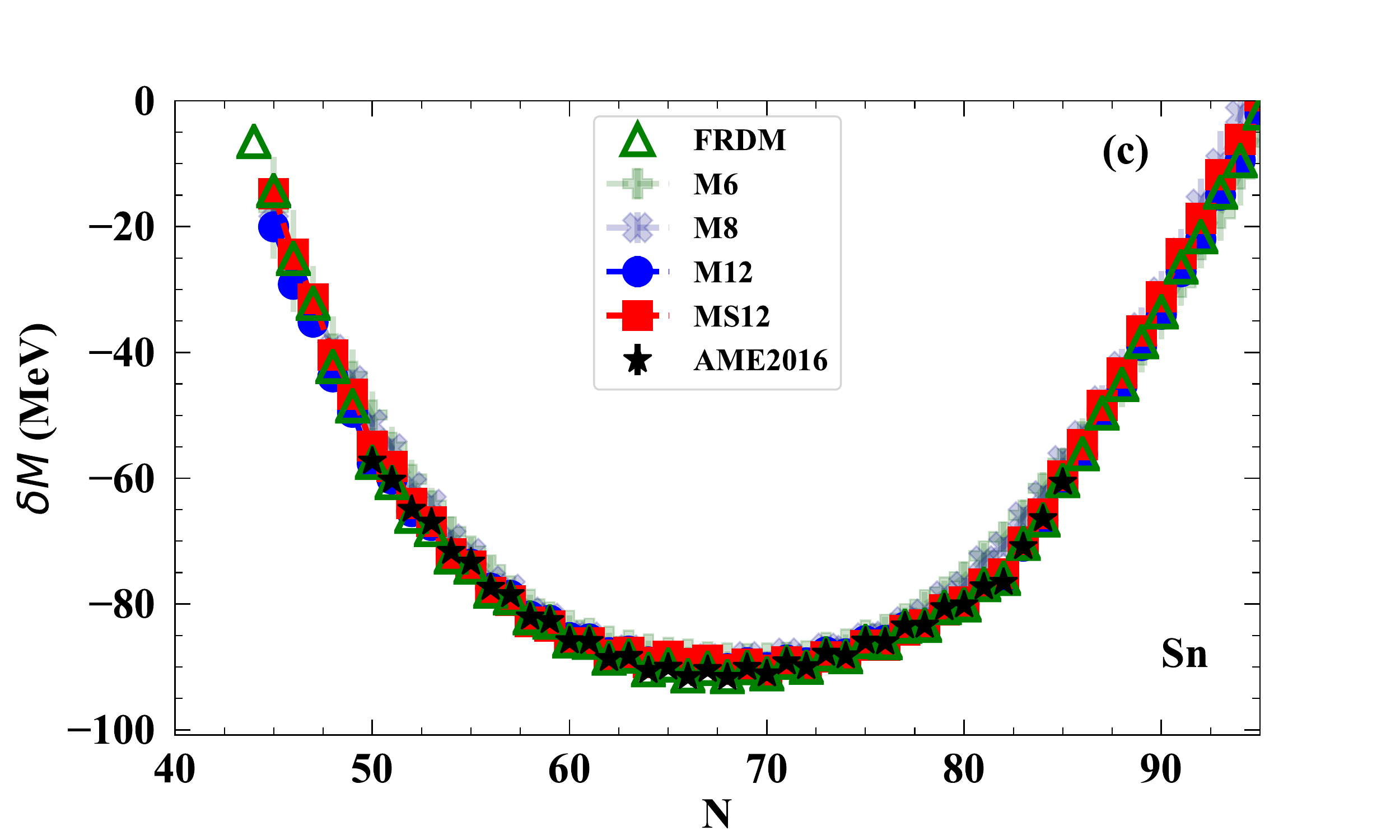} & \includegraphics[width=0.5\textwidth]{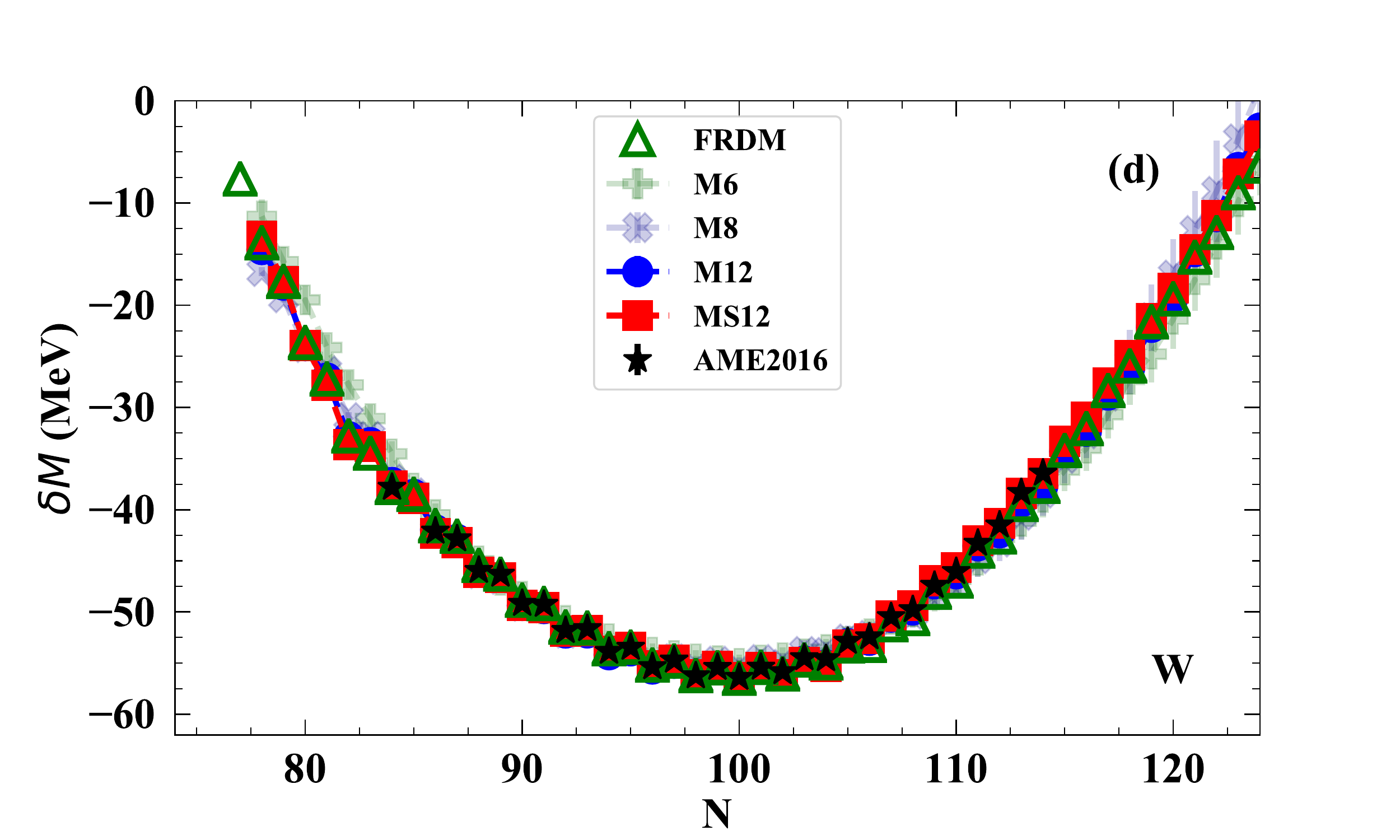} \\
\includegraphics[width=0.5\textwidth]{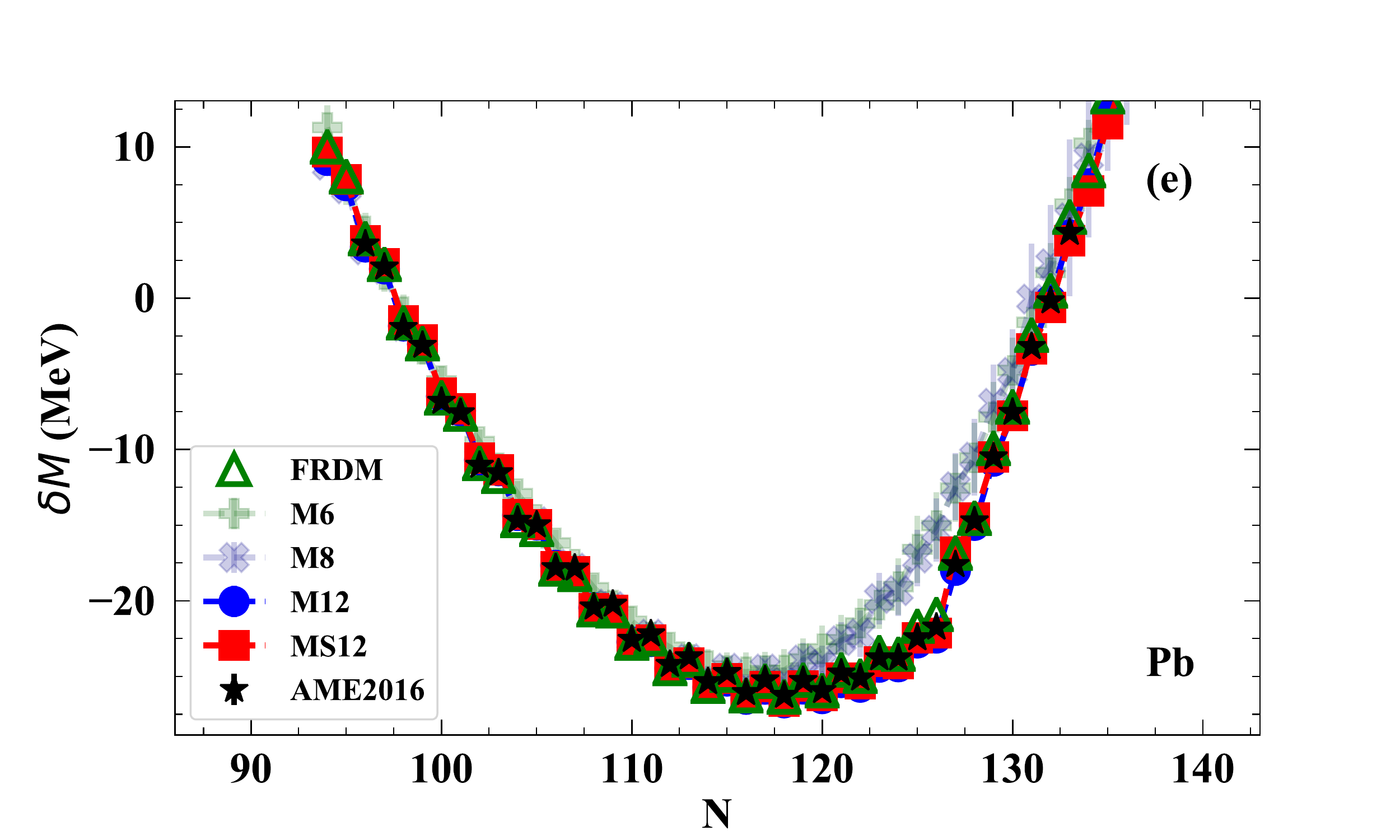} & \includegraphics[width=0.5\textwidth]{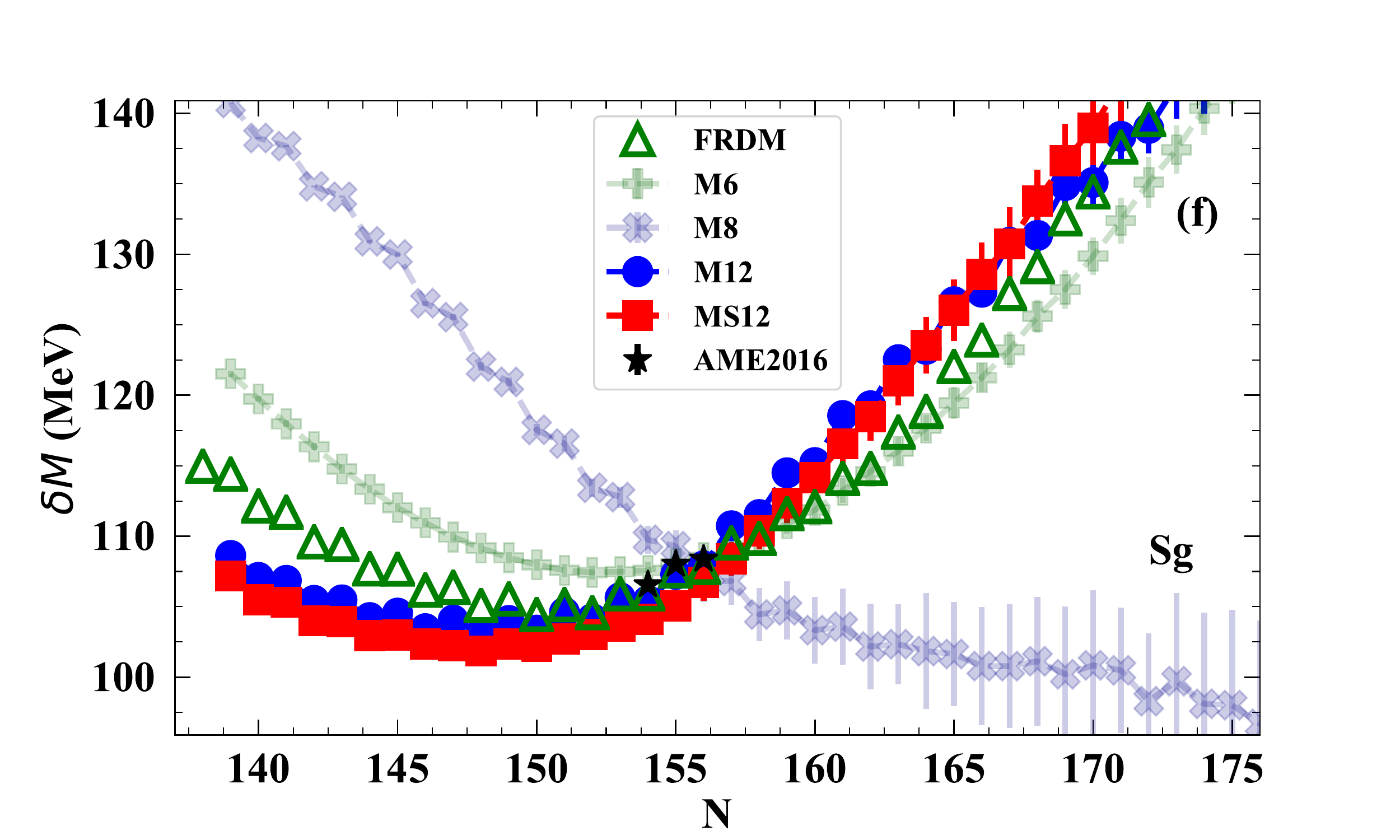} \\
\end{tabular}
\caption{Comparison of several MDN calculations--\emph{M6} (green pluses), \emph{M8} (dark blue x-es), \emph{M12} (blue circles), \emph{MS12} (red squares)--to the AME2016 (black stars) and FRDM2012 (green open triangles) for selected mass chains:  (a) Fe, (b) Mo, (c) Sn, (d) W, (e) Pb, and (f) Sg.}
\label{fig:massChains}
\end{figure*}

Along with reducing the $\sigma_\mathrm{RMS}$ between the MDN predictions and the AME2016, the second reason for including more features in the input space is to improve the predictions in the extrapolated region where experimental measurements have not been made or are recent enough to have not been included in the AME2016.  To look at the MDN predictions, we show the mass differences, $\delta M$, along a few select mass chains in Fig. \ref{fig:massChains}, (a) Fe, (b) Mo, (c) Sn, (d) W, (e) Pb, and (f) Sg.  We note that these mass chains span nearly the entire range of elements included in the training set, from Z=26 to Z=106, and are representative of the nuclear chart for $A\ge 20$.  For each mass chain, we plot four of the MDN calculations, \emph{M6}, \emph{M8}, \emph{M12}, and \emph{MS12}, along with the AME2016 to compare to the data, and FRDM2012 to show reasonable values for the extrapolated mass differences.  Here, we can see in more detail the benefits of adding more information to the feature space of the training set.  Going from \emph{M6} to \emph{M8}--which is particularly clear in panel (f) due to the poor predictive power of the MDN for these two training sets at higher Z values--we see that there is no odd-even staggering along the mass chain for \emph{M6}, but that this feature appears in \emph{M8} when a feature was included for whether the given nucleus has an odd or even number of neutrons/protons, as is seen in the reproduction of the training set as well.  In addition, we see that the odd-even staggering is present even for the extrapolated mass predictions, showing that the MDN algorithm has learned that the odd-even staggering is an important feature for predicting the masses.  Incorporating these additional features into the input space, even when the overall trend in the predictions is incorrect, is crucial for reproducing the physics trends in the predicted MDN values.

In the difference between \emph{M8} and \emph{M12}/\emph{MS12}, we also see the importance of including information about the magic numbers/shell closures.  It is particularly clear in Fig. \ref{fig:massChains}(e) (but also noticeable in the remaining panels) that the masses around the shell closures are better reproduced when information about this the magic numbers is included in the feature space.  Instead of smoothly following a parabolic shape (with or without corrections for the even-odd staggering), the larger gaps between $\delta M$ values at the shell closures are see in the mass chains.  In addition, one of the greatest benefits of including more features in the training set is to improve and stabilize the extrapolations, particularly noticeable in Fig. \ref{fig:massChains}(f), where only three masses from the chain are available to be included in the training set.  For the \emph{M8} training set, the extrapolated values of $\delta M$ do not even follow physical trends, simply decreasing as $N$ increases, instead of following a parabolic shape.  Interestingly, the \emph{M6} feature set provides a better prediction for this mass chain than the \emph{M8}.  However, this discrepancy is due more to the instability of the MDN predictions when the feature space is small (MDN calculations with similar values for $\sigma_{RMS}$ compared to the AME2016 can lead to widely discrepant mass predictions in the extrapolated region) rather than the predictions truly becoming worse as the feature space is increased, as will be discussed further in Sec. \ref{sec:robustness}.

\subsection{Reproducibility of the MDN predictions}
\label{sec:robustness}

\begin{figure}
    \centering
    \includegraphics[width=0.5\textwidth]{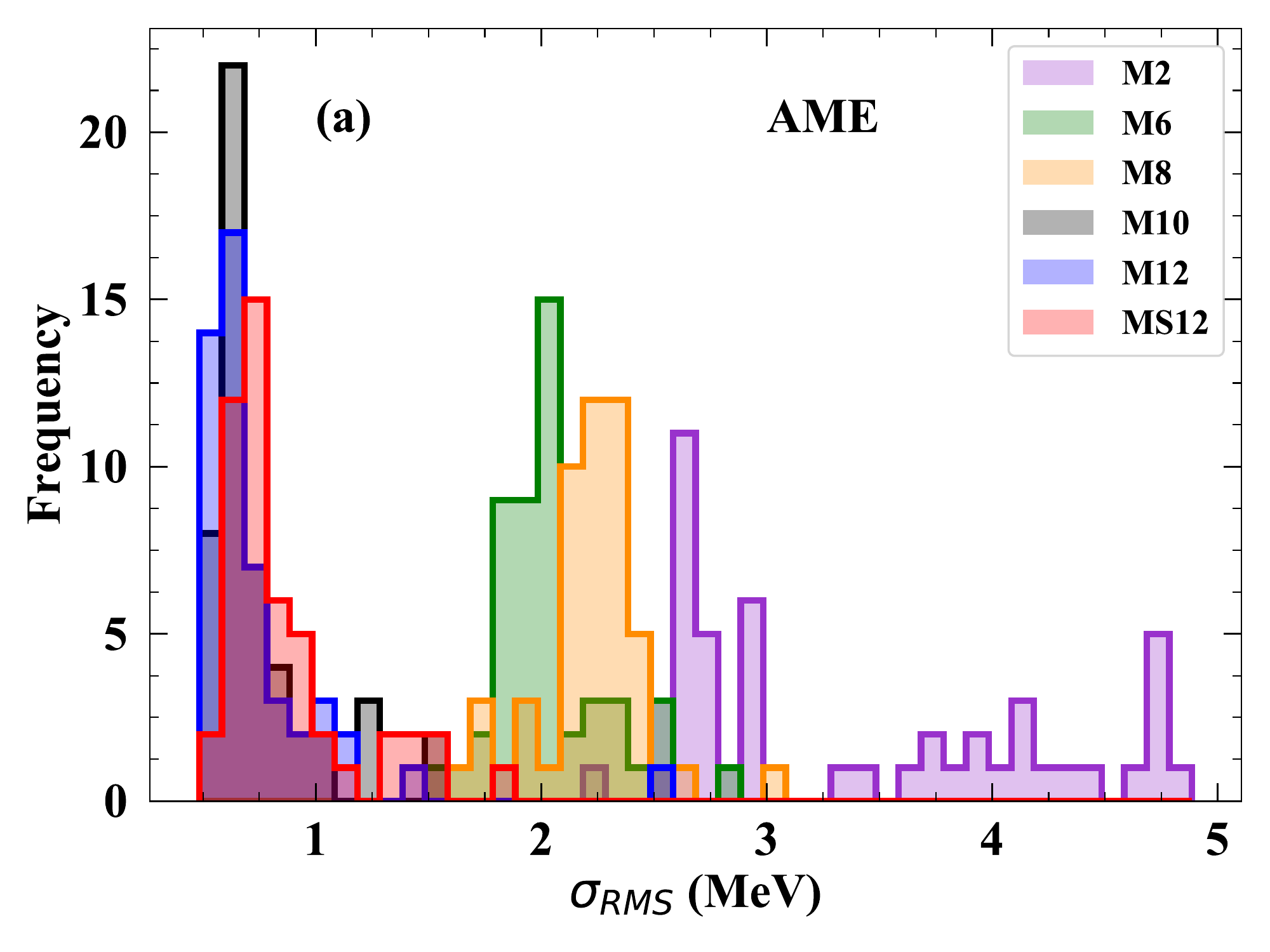} \\
    \includegraphics[width=0.5\textwidth]{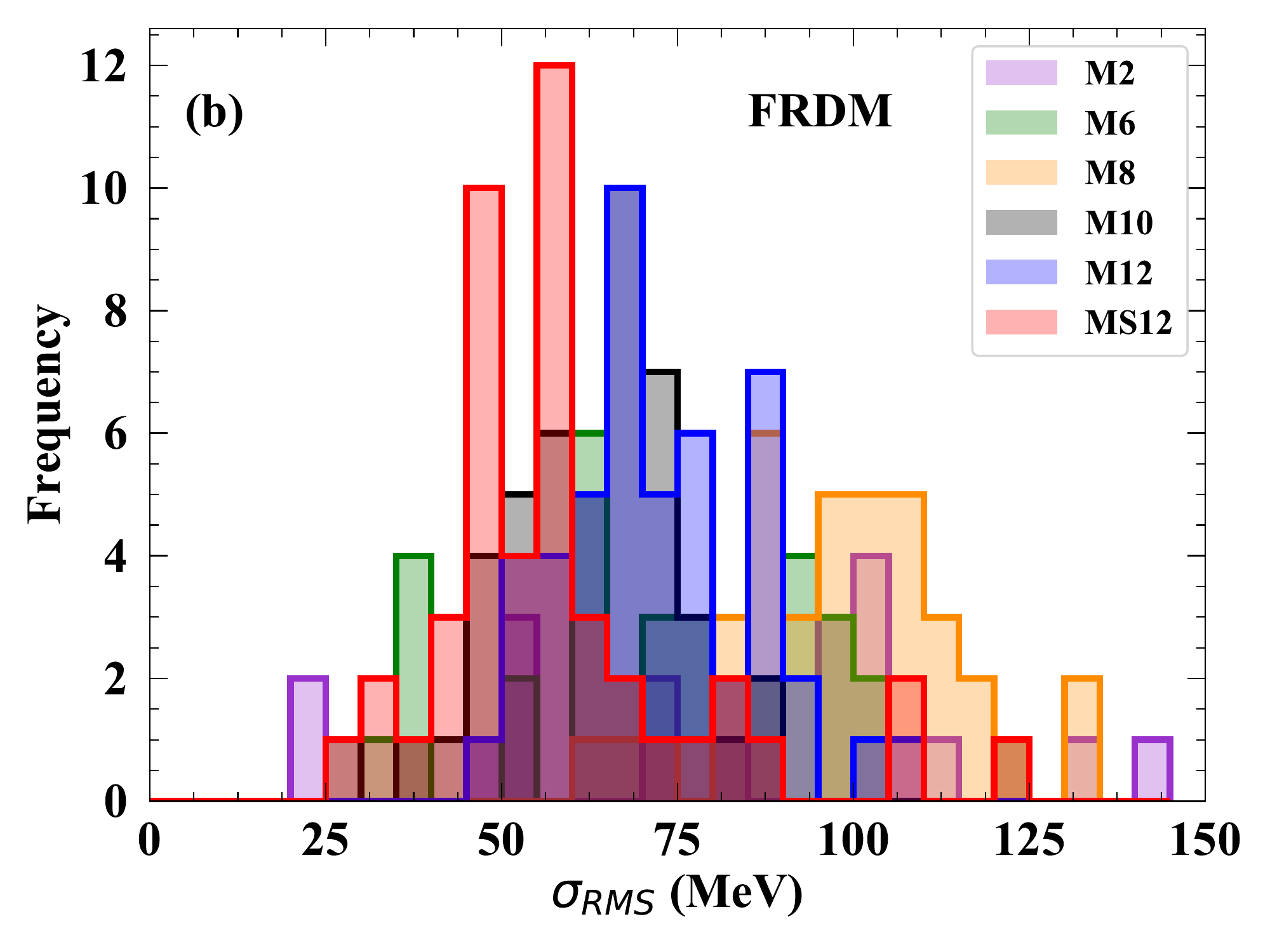}
    \caption{Distribution of $\sigma_\mathrm{RMS}$ values for each of the six MDN model predictions, in (a) the training region, compared to the AME, and (b) in the extrapolated region, compared to FRDM. }
    \label{fig:reproduceMass}
\end{figure}

We run the MDN training and prediction for each model 50 times to understand the changes in the reproducibility of the MDN algorithm as the size of the feature space is increased.  In Fig. \ref{fig:reproduceMass}, we show the distribution of the $\sigma_\mathrm{RMS}$ values for $\delta M$ (in MeV) for each of the models, compared to (a) the AME (includes only the masses in the training set) and (b) FRDM (includes the masses in the training set and the extrapolated region).  In Fig. \ref{fig:reproduceMass}(a), models $M10$, $M12$, and $MS12$ have very similar values for $\sigma_{RMS}$, as can be seen in Table \ref{tab:ResultsOverview} for the single calculation analysis.  In addition, the width of these three distributions are also fairly similar, with the tail of the $MS12$ distribution extending about 0.5 MeV farther than the distributions of $M10$ and $M12$.  (However, the number of counts in this extended tail is less than 5 out of 50.)  When reproducing the training set, the addition of features after the even-odd distinction--$M10$ and further feature additions--does not improve the reproduction of the training set, or the reproducibility across multiple MDN trainings.

It is in the extrapolation region that we see a marked improvement with the addition of the extended feature space, including training jointly on the mass excesses and one-neutron separation energies (again, only $MS12$ is shown, as the other feature spaces where both mass excess and neutron separation energy are included in the output space display the same trends between the $MN$ and $MSN$ feature spaces).  In Fig. \ref{fig:reproduceMass}(b), we show the distributions of the $\sigma_\mathrm{RMS}$ values for six of the MDN models with respect to FRDM, to show the overall reproducibility when the extrapolated region is taken into account.  On the scale of Fig. \ref{fig:reproduceMass}(b), each of the MDN models has a similar mean $\sigma_\mathrm{RMS}$ compared to FRDM.  However, the MDN models where more input features are included have a lower average $\sigma_\mathrm{RMS}$ across the 50 iterations of the training along with a narrower distribution.  (The narrower distribution of predictions with each increased feature space indicates that the prediction of $M8$ in Fig. \ref{fig:massChains}(f) are an outlier, but the training sets with a larger feature space lead to more reproducible predictions.)  

Here, we see also the benefit of including both the mass excess and one-neutron separation energy in the output of the train - instead of just the mass excess.  In particular, the distribution of results from the \emph{MS12} model has a lower average $\sigma_\mathrm{RMS}$ than the distribution from the \emph{M12} model.  
This result indicates, both, that having a more complete feature space for the network training improves the extrapolation of the observable of interest and that simultaneously predicting two related observables better constrains the extrapolation.  Even though the $\sigma_{RMS}$ does not decrease from \emph{M12} to \emph{MS12}, the extrapolated mass values are more consistent with one another.



\section{Conclusion}
\label{sec:conclusion}


In conclusion, we have used for the first time, the probabilistic machine learning technique, the Mixture Density Network (MDN), to directly learn the nuclear mass excesses using a series of features related to the properties of the nuclear mass models, with well-quantified uncertainties.
As the number of features were increased, the $\sigma_\mathrm{RMS}$ between the MDN predictions and the 2016 Atomic Mass Evaluation (AME) was greatly improved:  from nearly 4 MeV when only neutron and proton number were included in the feature set, down almost 500 keV when the full suite of features were included in the input space.  In addition, as more features were added to the input space, the extrapolation of masses beyond the AME was improved, in terms of the shape of each mass chain being more physical.  We also saw that simultaneously predicting both the mass excess and the one-neutron separation energy using the MDN further improved the quality of the extrapolated masses and the reproducibility of the MDN prediction.  Thus, including more physics information into the input space for the network training improved the performance of the network.

A logical next step for this work would be to test whether this expanded input feature space allows for fewer masses to be included in the training data set while keeping the $\sigma_\mathrm{RMS}$ of the mass excess compared to the AME close to or below 500 keV.  In addition, instead of including additional physics information into the input feature space, it is possible to include more physics directly into the neural network itself, either with further constraints built into the loss function or using a specially designed network architecture (e.g. recurrent neural networks are typically used to learn patterns in sequential data, and such a network design could possibly better capture the parabolic-like behavior of each mass chain).

Finally, one of the benefits of using the MDN is to provide realistic uncertainty estimates for each predicted values based on the uncertainty included in the training set.  To this end, the MDN could be used in a similar way as Gaussian Processes and Bayesian Neural Networks to perform Bayesian averaging of multiple mass models to predict the limits of stability, e.g.~\cite{Neufcourt2019,Neufcourt2020a}.  It would be interesting to compare the results from the MDN to predictions of other techniques, both in terms of the physics content and the quality of the uncertainty quantification.
 

\begin{acknowledgements}
	This work was performed under the auspice of the U.S. Department of Energy by Los Alamos National Laboratory under Contract 89233218CNA000001.  We gratefully acknowledge the support of the U.S. Department of Energy through the LANL/LDRD Program and the Center for Non Linear Studies.  
\end{acknowledgements}




\bibliography{Masses}

\end{document}